\documentclass[aps,pra,twocolumn,superscriptaddress]{revtex4}

\usepackage[utf8]{inputenc}  
\usepackage[pdftex]{graphicx}
\usepackage{amsmath,amsfonts,amssymb}

\newcommand{\rmc}{{\mathrm{c}}}
\newcommand{\rmd}{{\mathrm{d}}}
\newcommand{\rme}{{\mathrm{e}}}
\newcommand{\rmg}{{\mathrm{g}}}
\newcommand{\rmi}{{\mathrm{i}}}

\newcommand{\dB}{{\mathrm{dB}}}

\newcommand{\bfE}{{\mathbf{E}}}
\newcommand{\bfF}{{\mathbf{F}}}
\newcommand{\bfH}{{\mathbf{H}}}
\newcommand{\bfJ}{{\mathbf{J}}}
\newcommand{\bfe}{{\mathbf{e}}}

\newcommand{\bfr}{{\mathbf{r}}}
\newcommand{\bfv}{{\mathbf{v}}}

\newcommand{\cS}{{\mathcal S}}
\newcommand{\cV}{{\mathcal V}}

\newcommand{\ZZ}{{\mathbb Z}}

\begin{document}

\title{On frequency and time domain models of traveling wave tubes}

\author{St\'ephane~Th\'eveny}
\affiliation{Thales Electron Devices, 
             rue Lat\'eco\`ere, 2, FR-78140 V\'elizy.}%
\affiliation{Aix-Marseilles university, 
             UMR~7345 CNRS, PIIM, 
             \'equipe turbulence plasma, case 322 campus Saint-J\'er\^ome, 
             av.~esc.~Normandie-Niemen, FR-13397 Marseille cedex 20.}%
\author{Fr\'ed\'eric~Andr\'e}
\email{frederic.andre [at] thalesgroup.com} 
\affiliation{Thales Electron Devices, 
             rue Lat\'eco\`ere, 2, FR-78140 V\'elizy.}%
\author{Yves~Elskens}
\email{yves.elskens [at] univ-amu.fr}
%
\affiliation{Aix-Marseilles university, 
             UMR~7345 CNRS, PIIM, 
             \'equipe turbulence plasma, case 322 campus Saint-J\'er\^ome, 
             av.~esc.~Normandie-Niemen, FR-13397 Marseille cedex 20.}%


\begin{abstract}
We discuss the envelope modulation assumption of frequency-domain models of traveling wave tubes (TWTs) 
and test its consistency with the Maxwell equations. 
We compare the predictions of usual frequency-domain models 
with those of a new time domain model of the TWT. 
  52.35.Fp (Plasma: electrostatic waves and oscillations) \newline %
  52.40.Mj (particle beam interaction in plasmas) \newline %
  52.20.Dq (particle orbits) 
\end{abstract}
\maketitle

%
%

%

%
%
%
%

\section{Introduction}

{T}{wo approaches} are usually applied in describing 
the dynamics in a traveling wave tube (TWT) : 
(i)~the time domain approach makes no assumption on the TWT working frequency,
nor even on the shape of the signal wave, while 
(ii)~the frequency domain approach implies that all variables of interest 
depend on time like $\rme^{\rmi \omega t}$. 
So-called multi-frequency models can accommodate several frequencies, 
provided they are multiples of a fundamental frequency. 
This hampers the prediction of non-harmonic instabilities, as may result from nonlinearities or defects. 

Current simulations for TWT design rely mainly on frequency domain models 
because of their fast numerical execution. 
However, the current need for higher power and gain brings these models to the limits of their reliability. 
Therefore, we develop a family of time domain models \cite{An13,An13IVEC,An15,Be11}
inspired by \cite{Ku80,Ry07}, 
in order to provide more compact, accurate and complete descriptions of such regimes. 
This paper compares this time-domain approach and frequency models, as well as the foundations of the latter.

In section \ref{SecMaxwell}, we revisit a central assumption of
industrial frequency models
\cite{An97,Li02,Li09,Wa99} and discuss its consistency with Maxwell
equations when applied to three-dimensional geometry.  Then, we
rederive the basic equation for the modulation amplitude in frequency
models.  In section \ref{SecComp1}, we compare frequency-domain models
to assess the importance of the selection of modes which must be made
in order to run the model.  In section \ref{SecDimo}, we recall the
principles of our discrete model, which we compare to frequency-domain
models in section \ref{SecComp2} to assess their mutual consistency.

\section{Consistency of frequency models}
\label{SecMaxwell}

\begin{figure}[t]
      \centering
      \includegraphics[width=\columnwidth]{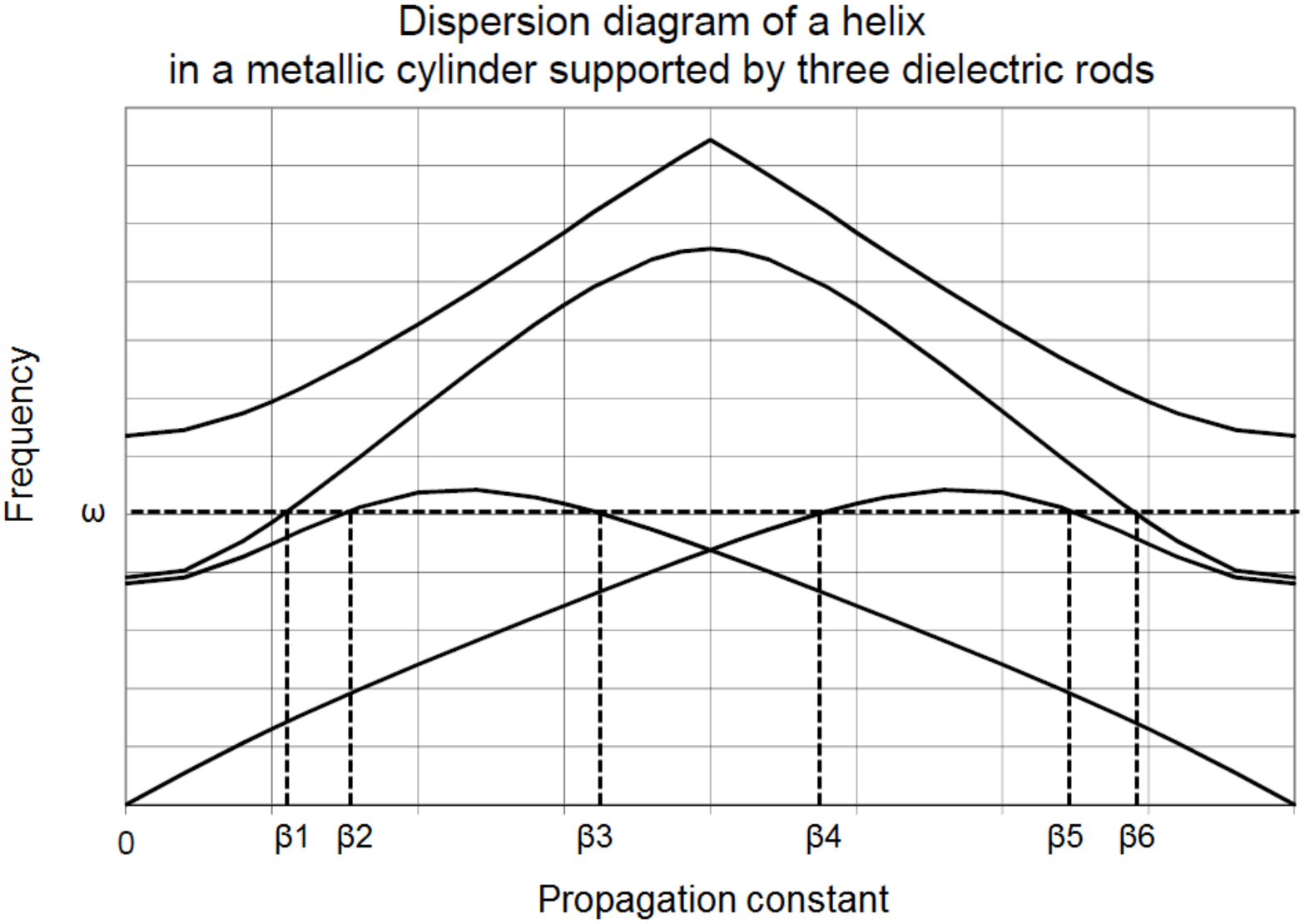}
      \caption{Typical dispersion diagram of a helix supported by
        dielectric rods inside a metallic cylinder. The helix can
        propagate six distinct modes at the chosen frequency.}
      \label{fig:HelixDispRel}
\end{figure}

Frequency domain models used in the industry share a common
representation of the electromagnetic field interacting with the
electron beam. The ``hot'' field at frequency $\omega$ is simply the
sum of ``beamless'' modes modulated by an envelope factor function of the
axial position $z$ only. One \cite{Wa99} or several \cite{An97,Li09}
modes can be used in this expansion. The term ``mode'' refers to the
different propagating modes existing in the delay line at the
considered frequency $\omega$, each mode $m$ having a real propagating
constant $\beta_m$ as shown in
Fig.~\ref{fig:HelixDispRel}. Consequently, we write in harmonic form
\begin{eqnarray}
  \bfE_{\rmc} (\bfr)
  &=&
  \sum_{m} C_{m}(z) \bfE_{m} (\bfr) ,
  \label{ApproxEnvE}
  \\
  \bfH_{\rmc} (\bfr)
  &=& 
   \sum_{m} C_{m}(z) \bfH_{m} (\bfr) .
  \label{ApproxEnvH}
\end{eqnarray}
The envelope factors are the $C_{m}$'s, and $z$ is the longitudinal
coordinate along the tube axis, with unit vector $\bfe_z$. In this
paper, we call such models Cold Wave Amplification Models (CoWAMs).
The real electric field is $\frac{1}{2}\Re \left( \bfE_\rmc(\bfr)
\rme^{\rmi \omega t}\right)$ when only one frequency is
considered. The case of several frequencies, in particular harmonics,
can be treated by summing the expansions at each considered
frequency. Hereafter, only one frequency is considered.

By definition, the cold fields $\bfE_{m}$ and $\bfH_{m}$ solve the
homogeneous Maxwell equations in harmonic form, viz.\ the Helmholtz
equation,
\begin{eqnarray}
  \nabla \wedge \bfE_m 
  & = & 
  - \rmi \omega \mu_{0} \bfH_m  ,
  \label{Mxw1}
  \\
  \nabla \wedge \bfH_m 
  & = & 
  \rmi \omega \epsilon_0 \bfE_m  ,
  \label{Mxw2}
\end{eqnarray}
with appropriate boundary conditions.   

In presence of a beam, the physical fields $\bfE (\bfr)$ and $\bfH (\bfr)$ 
must satisfy the full Maxwell equations, in particular Maxwell-Faraday for $\nabla \wedge \bfE$. 
The electric field then reads 
\begin{equation}
  \bfE (\bfr) = \bfE_{\mathrm{sc}}(\bfr) + \bfE_{\rmc} (\bfr),
\label{Esum}
\end{equation}
where the space-charge field $\bfE_{\mathrm{sc}}$ is a gradient.  Then
the curl of the full field $\bfE (\bfr)\rme^{\rmi \omega t}$ is only
the curl of the circuit field (\ref{ApproxEnvE}),
\begin{eqnarray}
  & & 
  \nabla \wedge \bfE(\bfr) 
  \nonumber \\
  & = & 
  \nabla \wedge \sum_m C_m(z) \bfE_m (\bfr)
  \nonumber \\
  & = & 
  \sum_m  \left( \frac{\rmd C_m}{\rmd z} \bfe_z \wedge \bfE_m (\bfr) 
  + C_m(z) \nabla \wedge \bfE_m (\bfr) \right)
  \nonumber \\
  & = &
  \sum_m  \left( \frac{\rmd C_m}{\rmd z} \bfe_z \wedge \bfE_m (\bfr) 
  - \rmi C_m(z) \, \omega_m \mu_0 \bfH_m (\bfr) \right) \!\!,
  \label{MxwEnv1}
\end{eqnarray}
where the second equality follows from a vector analysis identity 
and the third equality results from  (\ref{Mxw1}). 
Finally, recalling (\ref{ApproxEnvH}) yields
\begin{equation}
  \nabla \wedge \bfE_{\rmc} (\bfr, t) + \mu_0 \partial_t \bfH_{\rmc} (\bfr, t)
  = 
  \sum_m \frac{\rmd C_m}{\rmd z} \, \bfe_z \wedge \bfE_m (\bfr) \, \rme^{\rmi \omega t} .
\label{MxwEnv2}
\end{equation}
The Maxwell-Faraday equation states that the left-hand side must vanish. 
Therefore, the envelope form (\ref{ApproxEnvE})-(\ref{ApproxEnvH}) is consistent with the Maxwell equations 
only if the right hand side vanishes. If the electric field of each
mode is purely longitudinal, this condition is met automatically.

However, generically the electric field is not purely longitudinal
through the whole cross section of the structure. Otherwise the
Poynting vector would be purely transverse and the considered mode
would carry no power.  This precludes the possibility to use a single
mode in the expansion (\ref{ApproxEnvE}). Indeed, the only possibility
to satisfy the Maxwell-Faraday equation with a single cold mode is
that $\partial_z C_1=0$, i.e. that no amplification occurs. In
practice, $C_1$ is slowly varying, so that $\partial_z C_1$ remains low
enough for a good approximation.

The expansion on a single mode violates the Maxwell equations, so the
question is now to determine whether they can be satisfied by an
expansion on several modes.

To this end, consider the Maxwell-Amp\`ere equation  
\begin{equation}
  \nabla \wedge \bfH
  =
  \bfJ (\bfr, t) + \epsilon_{0} \partial_t \bfE
\label{MxwAmp1}
\end{equation}
and recall that $\bfH = \bfH_{\rmc}$.
Again, substituting (\ref{ApproxEnvE})-(\ref{ApproxEnvH})-(\ref{Esum}) 
and using the same vector identity yields 
\begin{eqnarray}
  \sum_{m} \frac{\rmd C_{m}}{\rmd z} \bfe_{z} \wedge \bfH_{m} \, \rme^{\rmi \omega t}
  = 
  \bfJ + \epsilon_{0} \partial_t \bfE_{\mathrm{sc}} .
\label{MxwAmp2}
\end{eqnarray}
Denote by $\tilde \bfJ_\omega (\bfr)$ the Fourier component of $\bfJ + \epsilon_{0} \partial_t \bfE_{\mathrm{sc}}$. 
Multiplying (\ref{MxwAmp2}) with $\bfE_n^*$ and using a vector identity obtains
\begin{eqnarray}
  \sum_{m} \frac{\rmd C_{m}}{\rmd z} \left(\bfH_{m} \wedge \bfE_{n}^{*} \right) \cdot \bfe_{z}
  = \tilde \bfJ_\omega \cdot \bfE_n^* . 
\label{Poy1}
\end{eqnarray}
Similarly, dot-multiplying (\ref{MxwEnv2}) with $\bfH_n^*$ and using a vector identity yields 
\begin{eqnarray}
  0  
  & = &  
  \sum_m \frac{\rmd C_m}{\rmd z} ( \bfE_m (\bfr) \wedge \bfH_n^* (\bfr) ) \cdot \bfe_z 
\label{Poy2}
\end{eqnarray}
at every point $\bfr$. 

Subtracting (\ref{Poy2}) from (\ref{Poy1})
and integrating over a transverse, planar section $\cS$ yields
\begin{eqnarray}
  &&
  \sum_{m} \frac{\rmd C_{m}}{\rmd z} \int_{\cS} 
      \left(\bfH_{m} \wedge \bfE_{n}^{*} - \bfE_{m} \wedge \bfH_{n}^{*}\right)
      \cdot \bfe_z \, \rmd^2 \bfr
  \nonumber \\
  & = & 
  \int_{\cS} \tilde \bfJ_\omega \cdot \bfE_n^* \, \rmd^2 \bfr  .
  \label{Puiss}
\end{eqnarray}
For a periodic structure (as considered in 
{\textsc{mvtrad}} \cite{Wa99}, {\textsc{christine}} \cite{An97} and {\textsc{bwis}} \cite{Li09}),
the integral in the left hand side vanishes for $m \neq n$, 
thanks to the mode orthogonality relation \cite{Ch07} 
\begin{eqnarray}
  \frac{1}{4} \int_{\cS} \left(\bfE_m \wedge \bfH_n^* + 
                                     \bfE_n^* \wedge \bfH_m \right) \cdot \bfe_z
                 \,  \rmd^2 \bfr
  = \delta_{mn} \sigma_m P_m 
  \label{ortho}
\end{eqnarray}
where $P_m$ is the absolute value of the (longitudinal)
electromagnetic power flow in mode $m$ (incorporating the
normalization of the eigenfields $\bfE_m$ and $\bfH_m$) and
$\sigma_m = \pm 1$ according to the direction of the power
flow. Equation (\ref{Puiss}) so reduces to
\begin{equation}
  \frac{\rmd C_m}{\rmd z} 
  = 
  - \frac{\sigma_m}{4 P_m} \int_{\cS} \tilde \bfJ_\omega \cdot \bfE_m^* \, \rmd^2 \bfr  .
  \label{dCn1}
\end{equation}

These are necessary, but not sufficient, conditions to satisfy the
Maxwell equations. Equations~(\ref{dCn1}) completely determine the
amplitude of the wave inside the TWT, along with the dynamics of the
electrons. At this point, we can make an important remark on which
modes will be amplified or not. At the start of the amplification
process, all modes have small amplitude and the TWT operates in the
linear regime. We know that in this case the beam is carrying space
charge waves \cite{Lo60}. These waves have their own wave number which is given by
the beam velocity $v_0$ in a first order approximation,
$\beta_0=\omega/v_0$. Therefore, only modes with $\beta_m$ close to
$\beta_0$ will grow significantly in equation~(\ref{dCn1}), because
otherwise the right hand side is the integral of an oscillating
function. This will be illustrated in the next section with a
numerical example.

Now, we compare the relative importance of these waves in term of power
by calculating the total power flow from the Poynting vector. The same
orthogonality theorem gives (with c.c.\ denoting complex conjugate)
\begin{equation}
  P = \frac{1}{2} \int_\cS \bfE_\rmc \wedge \bfH_\rmc^*\, \rmd^2 \bfr + \textrm{c.c.}
     = \frac{1}{2} \sum_m |C_m|^2 P_m ,
\label{eq:power}
\end{equation}
i.e., the total power results from the power carried by each
individual mode. Introducing the mode amplitude $a_m = C_m  \sqrt{P_m}$ \cite{Lo60},
the power of each mode is simply $|a_m|^2$ and
equation~(\ref{eq:power}) becomes
\begin{equation}
  \frac{\rmd a_m}{\rmd z} 
  = 
  - \frac{\sigma_m}{2\sqrt{2}}\beta_m \sqrt{Z_{\rmc m}} \int_{\cS} \tilde \bfJ_\omega \cdot {\mathcal{E}}_m^* \, \rmd^2 \bfr  .
  \label{da_m}
\end{equation}
where $Z_{\rmc m}$ is the coupling impedance of mode $m$ at the working
frequency and ${\mathcal{E}}_m$ is the electric field of that mode divided by
its amplitude.  With this final form of the interaction equation, we
can see that modes with large coupling impedance will rapidly dominate
over modes of low coupling impedance as we move toward the TWT output.

To conclude this part, we are left with a contradiction. We have just
seen that only the modes synchronized with the beam can grow, as
confirmed by dedicated experiments \cite{Do05}. In the usual situation
of a practical TWT, only one mode satisfies this synchronization
condition. Therefore, this mode alone rapidly dominates in the
amplification process. But we have also seen that the Maxwell
equations cannot be satisfied with a single mode. It follows that the
envelope model (CoWAM) can only approximate the physics. How accurate
this approximation is remains difficult to assess. One motivation to
develop the discrete model, beside its time domain capabilities, is to
lift these theoretical approximations. Based on our experience
however, we expect that the approximation is minor, at least inside
common ranges of parameters used for practical devices. This fact will
also be illustrated hereafter with a numerical example.

\section{Relevance of backward wave in CoWAM}
\label{SecComp1}

In this numerical example, we assess the relative importance of the
forward and backward modes for a standard helix tube. Such a tube
comprises attenuating sections, where the fields are partly absorbed~;
these attenuating sections are modeled in CoWAMs on adding a loss term
in (\ref{dCn1}) to give
\begin{equation}
  \frac{\rmd C_{1}}{\rmd z}
  =
  - \frac{\sigma_{1}}{4P_{1}} \int_{\cS} \bfE_{1}^{*}(z) \cdot \bfJ(z) \, \rmd^2 \bfr - \alpha'(z) C_1(z)
  \label{ResolC1}
\end{equation}
where the spatial loss rate $\alpha'$ is a function of position
accounting for propagation losses and localized attenuator. The
purpose of the present calculation is to compare the relative power of
the different modes radiated by a given modulated electron
beam. Consequently, the current density $\bfJ$ is an input generated
from a preliminary run of {\textsc{mvtrad}}. The same modulated
current density is used to calculate the radiation on the forward mode
alone or on the forward and backward modes together. The differences
in the resulting electric field acting on the electron trajectories is
not taken into account, i.e.\ these trajectories are frozen.

\begin{figure}[t]
  \centering
  \includegraphics[width=\columnwidth]{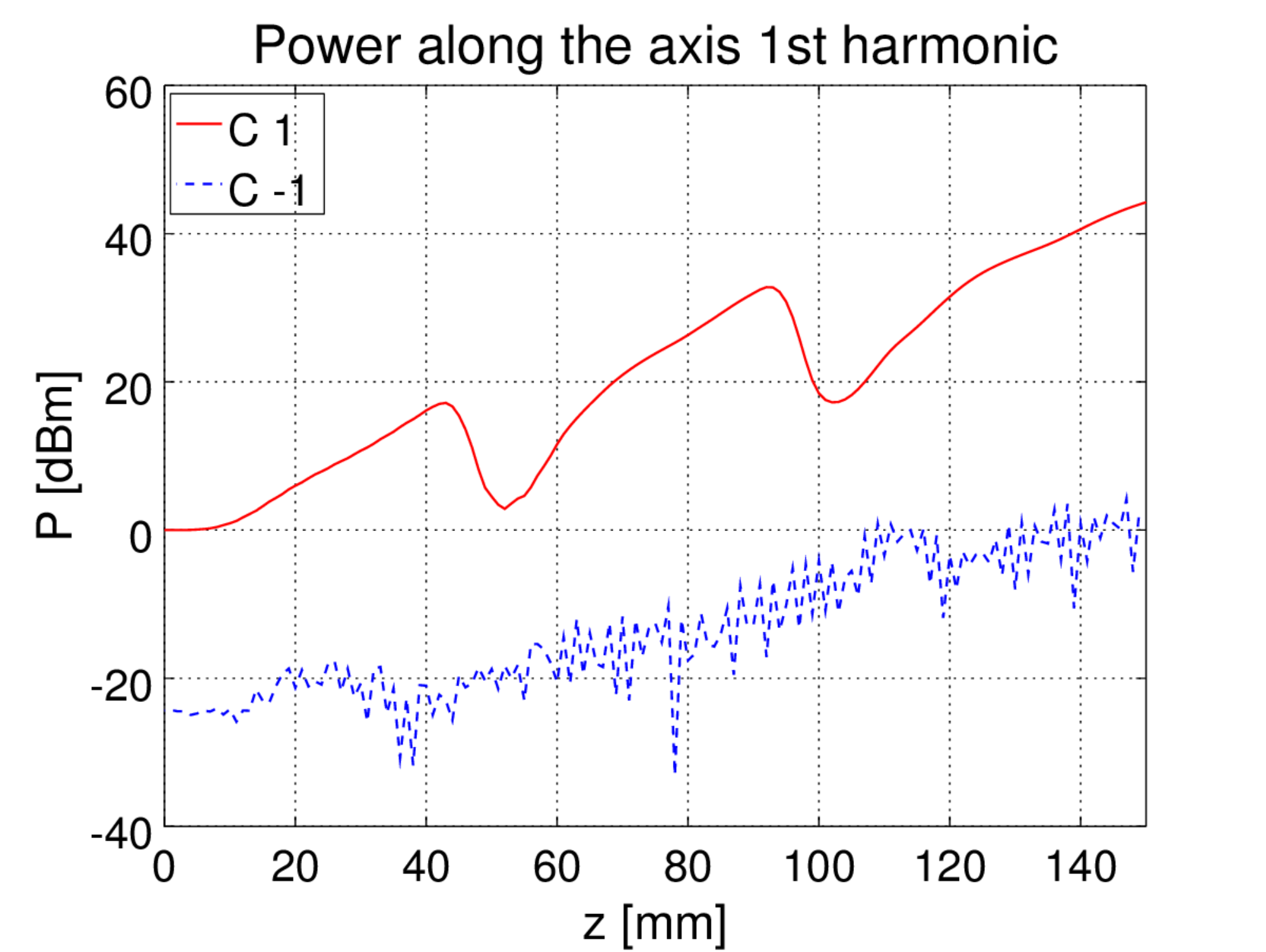}
  \caption{Power along the axis for the first harmonic.  Full line :
    contribution from coefficient $C_{+1}$.  Broken line :
    contribution from coefficient $C_{-1}$.  
    Parameters : $F = \omega / (2 \pi) = 11.5$~GHz, tube length 150~mm, 
    attenuating sections centered at 50~mm and 100~mm.}
      \label{fig:PowerCnsup}
\end{figure}
We first compare the power radiated in the model taking a single mode
into account ($m=1$) versus taking also the backward wave into account
($m = \pm 1$).  The full line on Fig.~\ref{fig:PowerCnsup} results
from integrating (\ref{ResolC1}) given a modulation current computed
with only one mode and computing the power gain as
\begin{eqnarray}
  P_{\dB}
  =
  10 \log_{10} \left(\frac{|C_{1}|^{2}}{C_{0}^{2}}\right)
\end{eqnarray}
where $C_0 = C_1(0)$ is the envelope amplitude at the TWT inlet. 

The broken line is obtained similarly for $m = - 1$, using also
{\textsc{mvtrad}} formally.  Indeed, though {\textsc{mvtrad}} does not
use the backward mode to compute the wave amplification,
(\ref{ResolC1}) can be integrated with the actual current density and
the eigenfield $\bfE_{-1}$, to check whether the beam modulations
might be resonant with the backward mode.  Fig.~\ref{fig:PowerCnsup}
shows that the backward mode is poorly coupled for this case, where
the phase shift per cell $\beta d$ is moderate~: 
near the outlet, the backward mode reaches 5~dB while the
direct mode culminates at 45~dB, and through the whole tube the
backward mode is always at least 20~dB weaker than the direct mode.

Quite obviously, the forward wave is significantly damped by the two
attenuating sections, inserted to avoid multiple reflexions in each
section. The attenuation of the backward wave is less visible because
its power is not building up, the synchronization condition being not
fulfilled. It results that the power radiated by the beam at any
location is not negligible compared to the incoming power carried by
the wave, including inside the attenuating section.

\begin{figure}[t]
      \centering
      \includegraphics[width=\columnwidth]{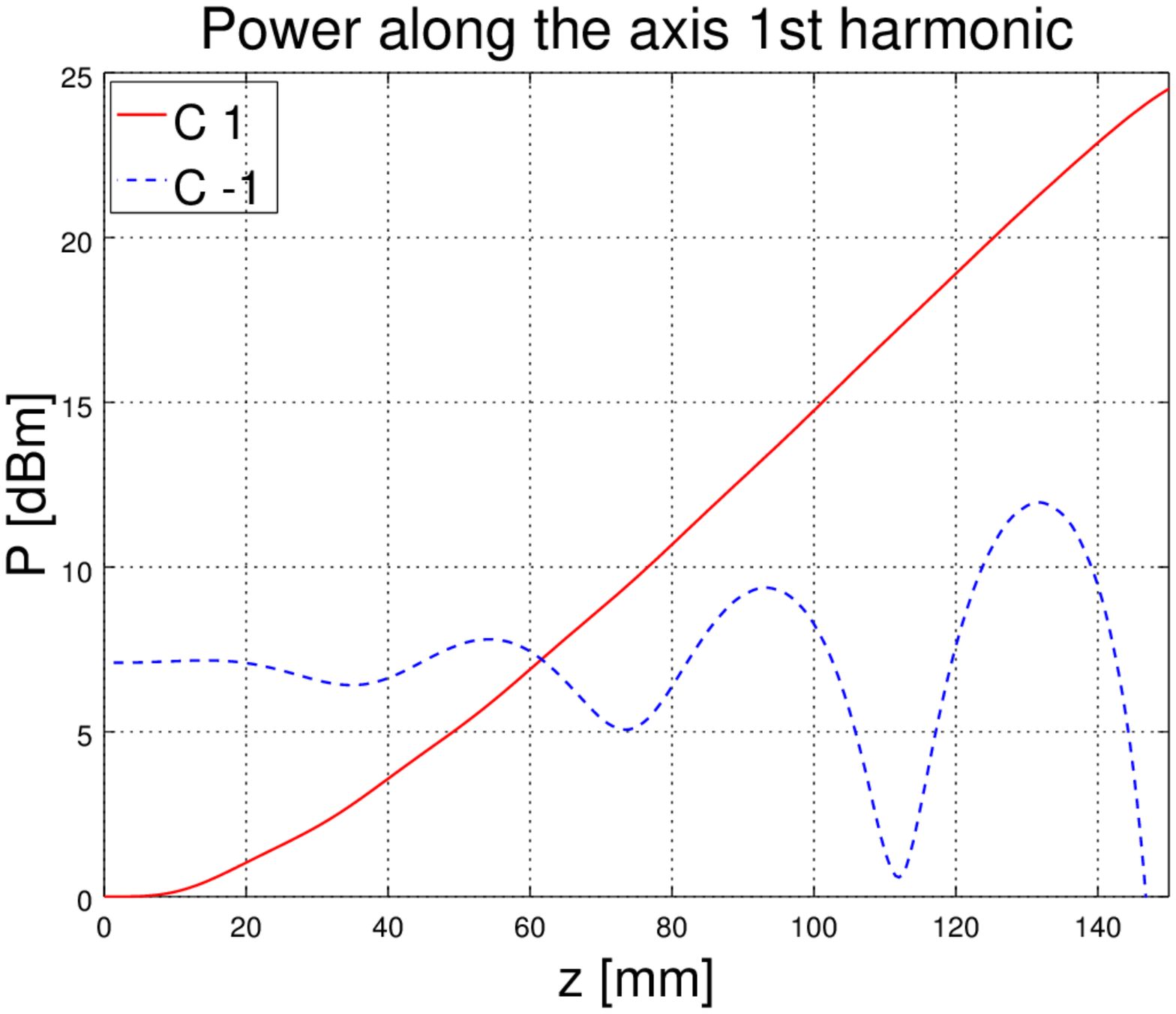}
 	 \caption{Same as Fig.~\ref{fig:PowerCnsup}, near the $\pi$ mode for
            $\beta = \pi/d$ 
            and no attenuation.}
    \label{fig:PowerCninf}
\end{figure}

The role of backward waves is more relevant when the TWT is operated
near the $\pi$ mode because they synchronize with the beam at this
operating frequency.  This is confirmed by our simulations. For this
simulation, we need a much higher frequency (out of the actual
operation band), at which the coupling impedance is much smaller so
that amplification occurs more slowly.  Near the $\pi$ mode, the
backward wave couples efficiently with the slow space charge wave of
the beam, generating the backward wave instability, where there is a
risk of developing an oscillation.  To stress the effect, our
simulation is run without attenuation, and shows that the backward
wave generated near the outlet of the TWT has an intensity exceeding
significantly the input of the direct wave (here by about 8~dBm at $z
= 0$). So, near the $\pi$ mode indeed, the CoWAM must incorporate the
backward wave, and a model like {\textsc{bwis}} will definitely be
more accurate than {\textsc{mvtrad}}.

However, a CoWAM is unable to predict an instability if it is not
foreseen by a genuine identification of the modes to be included in
the run.  Time-domain models will not suffer from this lack of
prescience.

\section{Discrete model}
\label{SecDimo}

In contrast with the previous models, our model {\textsc{dimo}} is
formulated in the time domain \cite{An15} and does not make use of the
same expansion on cold fields that has been critically reviewed
in section \ref{SecMaxwell}.  It implements the analysis presented in Ref.~\cite{An13}
and can describe any delay lines~: folded waveguides, helices, coupled
cavities\dots

We outline here the derivation of the discrete model which has been
presented in detail in the past.

To take advantage of the periodicity of the slow wave structure, one
first determines the solutions $\bfE_{s \beta} (\bfr)$, $\bfH_{s
  \beta} (\bfr)$ to the Helmholtz equation in a periodic cell (denoted
$\cV_0$, typically $0 \leq z \leq d$, with a given profile $(x,y) \in
\cS(z)$, e.g.\ $x^2 + y^2 \leq R^2$ for a cylinder), with the Floquet
boundary condition $\bfE_\beta (\bfr + d \bfe_z) = \rme^{-\rmi \beta
  d} \bfE_\beta (\bfr)$ for the solenoidal (or circuit) fields
$\bfE_\rmc$, $\bfH_\rmc$.  For each propagation constant $0 \leq \beta
< 2 \pi / d$, these solutions form a basis with eigenfrequencies
$\Omega_{s \beta}$.  Subscript $s$ labels different eigenmodes meeting
the same Floquet condition, i.e.\ the different bands (four of them
are displayed in Fig.~\ref{fig:HelixDispRel}).

Physical functions $G (\bfr, t)$ are expressed in terms of these eigenmodes
by first applying the Gel'fand $\beta$-transform \cite{Ku80}
\begin{eqnarray}
  G (\bfr + n d \bfe_z, t)
  & = &
  \frac{d}{2 \pi} \int_{0}^{2 \pi / d} G_{\beta} (\bfr, t) \, \rme^{- \rmi n \beta d} \, \rmd \beta ,
  \label{TransBetaI}
  \\
  G_{\beta} (\bfr, t) 
  & = &
  \sum_{n = - \infty}^{+ \infty} G (\bfr + n d \bfe_z, t) \, \rme^{\rmi n \beta d}   ,
  \label{TransBeta}
\end{eqnarray}
and expanding $\bfE_\beta$ and $\bfH_\beta$ on the Floquet divergence-free eigenbasis
of the slow wave structure,
\begin{eqnarray}
  \bfE_{\beta}(\bfr, t)
  & = &
  \sum_{s} V_{s \beta}(t) \bfE_{s \beta}(\bfr) - \nabla\phi_{\beta} ,
  \label{eq:bfE}
  \\
  \bfH_{\beta}(\bfr, t)
  & = &
  \rmi\sum_{s}I_{s \beta}(t)\bfH_{s \beta}(\bfr) ,
  \label{eq:bfH}
\end{eqnarray} 
where the first equation includes the irrotational space-charge field
contribution, described with the potential $\phi$, and the imaginary
unit is introduced so that final physical variables of the model are
purely real numbers. 

This expansion is the essential difference
between the discrete and the envelope models. The $\bfE_{s \beta}$ are
the (vector) eigenfunctions of the Helmholtz equation, therefore they clearly
constitute a basis on which any divergence-free vector field can be
expanded at any time. 
The discrete model makes no approximation in the function choice. 
On the contrary,
the expansion (\ref{ApproxEnvE})-(\ref{ApproxEnvH}) is performed on
the propagating modes at a given frequency which apparently do not
necessarily constitute a basis~: at a frequency below the cut-off of a waveguide,
for example, evanescent modes do exist although there are no
propagating modes on which they could be expanded. Even at a frequency
where propagation is possible, the modes with higher cut-off frequency
are still possible in the form of evanescent modes but cannot be
described by the modes propagating at this frequency.

The time-dependent coefficients $V_{s \beta}, I_{s \beta}$ should
generally not coincide (nor be merely proportional) as they will obey
their own coupled evolution equations, whereas the envelopes $C_m$ in
(\ref{ApproxEnvE})-(\ref{ApproxEnvH}) had to coincide to describe the
same modulation of cold fields in CoWAMs.

The Maxwell equations for the field propagation translate into 
evolution equations for the Floquet coefficients $V_{s \beta}, I_{s \beta}$ :
\begin{eqnarray}
  \dot{V}_{s \beta} + \Omega_{s \beta} I_{s \beta}
  &=&
  - \frac{1}{N_{s \beta}} \int_{\cV_{0}}\bfJ_{\beta} (\bfr, t) \cdot \bfE_{s \beta}^{*} (\bfr) \, \rmd^{3} \bfr  ,
  \label{eq:Vbd}
  \\
  \dot{I}_{s \beta} - \Omega_{s \beta}V_{s \beta}
  &=&
  0  ,
  \label{eq:Ibd}
\end{eqnarray}
where $N_{s \beta}$ is the electromagnetic energy of mode $(s, \beta)$ in a unit cell 
of the slow wave structure, 
and $\bfJ$ is the beam electric current density. 

Imposing a constant ratio $V_{s \beta} / I_{s \beta}$ in the discrete model (by analogy with the $C_m$'s)  
would be an extra condition, making the set of equations for the coefficients overdetermined.
Keeping linearly independent coefficients $V_{s \beta}, I_{s \beta}$ enables the discrete model 
to satisfy Maxwell equations, in contrast with CoWAMs. 

On introducing the field (closely related to the vector potential, see \cite{An13}) 
\begin{eqnarray}
  \bfF_{s, n} (\bfr) 
  & := & 
  \frac{1}{2\pi} \int_{0}^{2\pi} \frac{\bfE_{s \beta} (\bfr)}{N_{s \beta}} \rme^{- \rmi n \beta d} \, \rmd(\beta d) ,
\end{eqnarray}
these equations read in real space
\begin{eqnarray}
  \dot{V}_{sn} + \sum_{m}\Omega_{s,m}I_{s,n-m}
  &=&
  - \int_{\cV_{\ZZ}}\bfJ(\bfr, t) \cdot \bfF_{s,-n} (\bfr) \, \rmd^{3} \bfr   ,
  \nonumber \\ &&
  \label{Eq.Vsn}
  \\
  \dot{I}_{sn} - \sum_{m}\Omega_{s,m}V_{s,n-m}
  &=&
  0  ,
  \label{Eq.Isn}
\end{eqnarray}
where $\cV_{\ZZ}$ is the full extent of the slow wave structure. 

In real space and time variables, (\ref{eq:bfE})-(\ref{eq:bfH}) give the fields in terms of these coefficients as
\begin{eqnarray}
  \bfE (\bfr, t)
  &=&
  \sum_{s,n} V_{sn}(t) \, \bfE_{s,-n} (\bfr) - \nabla \phi(\bfr, t)  ,
  \label{EqRecEDIMO}
  \\
  \bfH (\bfr, t)
  &=&
  \rmi\sum_{s,n}I_{sn}(t) \, \bfH_{s,-n}(\bfr)  ,
  \label{EqRecHDIMO}
\end{eqnarray}
where $\phi(\bfr,t)$ is the beam space charge potential.

The detailed modeling of the slow wave structures lies (i)~in the frequency matrix $\Omega_s$,
which is typically a band matrix $\Omega_{s, n, m} = \Omega_{s, n-m}$, 
symmetric ($\Omega_{s, n, m} = \Omega_{s, m, n}$ by reciprocity condition), 
with a rather short range ($\Omega_{s, n-m} = 0$ if $| n - m | > p$ with, 
say, $p = 1$ for coupled cavities \cite{Ry07} and $p \sim 5$ for a helix),
and (ii)~in the explicit functions $\bfE_{s,n} (\bfr)$, $\bfF_{s,n} (\bfr)$, $\bfH_{s,n} (\bfr)$, 
whose construction involves the coupling impedances $Z_{s \beta}$ over the relevant bandwidth.
As a first approximation, one mode $s$ suffices to capture the physics of the TWT.
In contrast with the local picture of envelope modulation (\ref{ApproxEnvE})-(\ref{ApproxEnvH}), 
the fields $\bfE_{s, n}$ and $\bfF_{s, n}$ should not be viewed as local to a cell $n$ 
but rather may have quite long a range in terms of $n$ to express how the cell couples with the beam. 

The beam coupled with the wave is described by macro-electrons, 
with charge to mass ratio $- \eta = - | e | / m_\rme$ and mass $m$,
position $\bfr_k(t)$ and velocity $\bfv_k = \dot \bfr_k$, 
injected at cathode potential $- V_{\textrm{K}}$ 
at a constant rate to match the physical current $I_{\textrm{beam}}$ at the electron gun.
Their equation of motion then reads 
\begin{equation}
  \frac{\rmd}{\rmd t} [ (1 - | \bfv_k|^2 / c^2)^{-1/2} \bfv_k ]
  =
  - \eta [ \bfE (\bfr_k, t) + \mu_0 \bfv_k \wedge \bfH (\bfr_k, t) ] .
  \label{Eq.dvdt}
\end{equation}
The space-charge potential solves the Poisson equation with 
particles as sources \cite{Ro65} 
and boundary conditions fixed by the slow wave structure.

\section{Comparison of frequency and time domain models}
\label{SecComp2}

We now compare the cold wave amplification model in frequency domain
with the discrete model in time domain. Given the experimental
characteristics (dispersion diagram, coupling impedance) of the tube,
we interpolated the eigenmode electric field on the axis
$\bfE_{\beta}(z)$ and energy density $N_{\beta}$, and constructed the
matrix $\Omega_{n-m}$ and interaction field $\bfF_n(z)$ defined by the
Gel'fand transform.

Once again, the current modulation $\bfJ$ is obtained from a
preliminary \textsc{mvtrad} run and is injected in equation~(\ref{Eq.Vsn}). 
This procedure permits to compare the discrete
model and CoWAM with exactly the same current modulation. Indeed,
these two models differ only in how electromagnetic waves are
radiated from the beam, not in the dynamics of electrons given by (\ref{Eq.dvdt}).

\begin{figure}[t]
      \centering
      \includegraphics[width=\columnwidth]{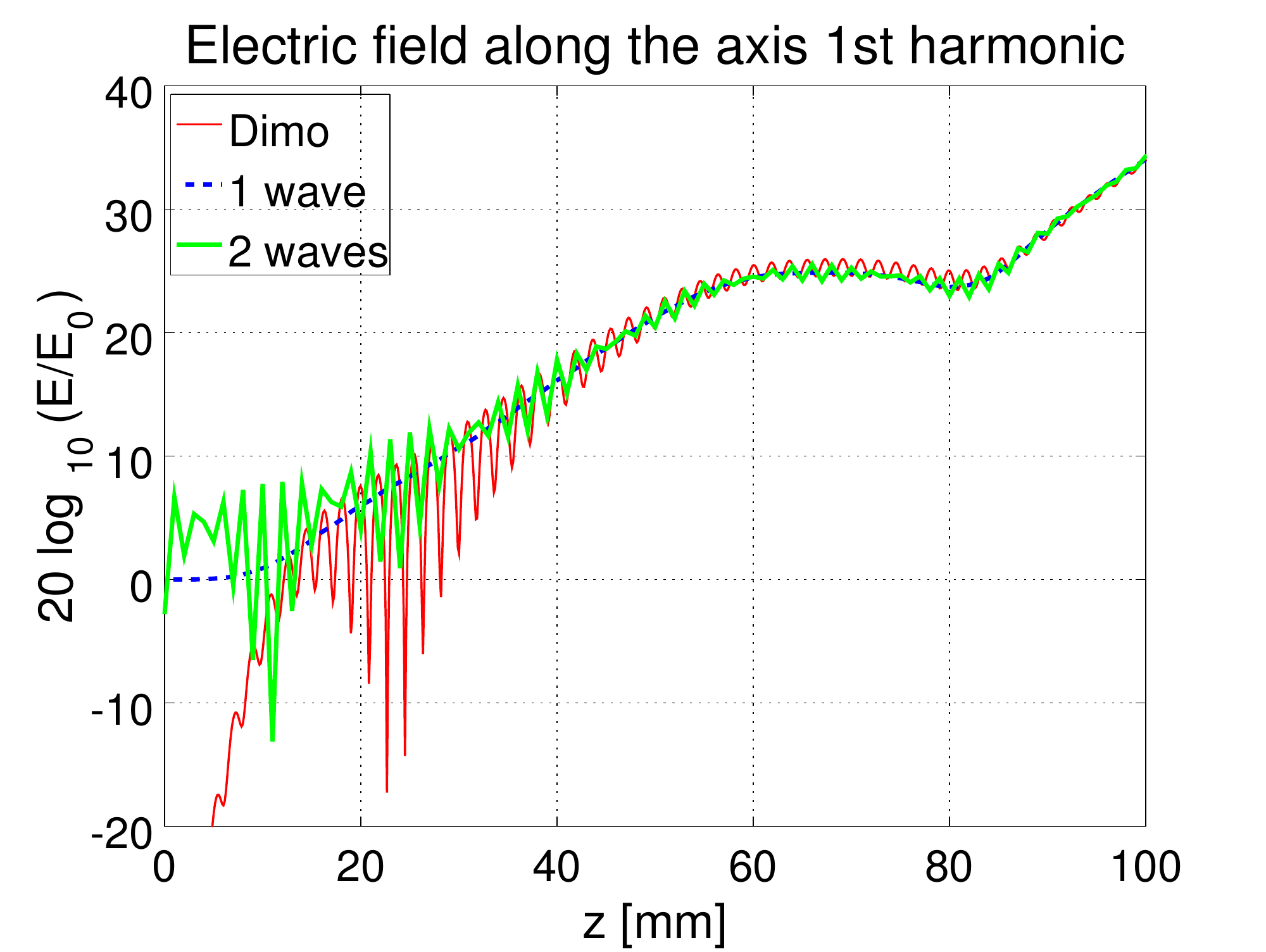}
 	  \caption{First harmonic of the electric field along the axis.
	    Thin solid red line~: {\textsc{dimo}}. 
	    Blue dots~: forward model {\textsc{mvtrad}}.
	    Thick solid green line~: forward and backward ({\textsc{bwis}}-like) model. 
	    Parameters~:
            $F = 11.5$~GHz, tube length 150~mm, no attenuation. 
          Ordinate scale shows the power rather than field amplitude. }
        \label{fig:ErrLogChpE}
\end{figure}

For {\textsc{mvtrad}} and the {\textsc{bwis}}-like model respectively, 
we solved (\ref{dCn1}) with $m=1$ and $m = \pm 1$,
as in section \ref{SecComp1}. 
All three models are run lossless (viz.\ we insert no attenuation). 

Fig.~\ref{fig:ErrLogChpE} displays the power in the first harmonic along the tube, normalized to the input power. 
All three models agree qualitatively, and rather well quantitatively, for the growth of the main harmonic. 
Because of their richer harmonic structure, {\textsc{bwis}} and {\textsc{dimo}} show more oscillations 
near the entrance to the tube, but these oscillations become negligible once the field becomes so intense 
that the first harmonic dominates. 
The end of the tube was not displayed on this figure, to zoom on the region where the contrast between models is larger. 
The shoulder near $z = 80$~mm is a saturation effect before the amplification resumes, 
and it is identically predicted by the three models, confirming their agreement. 

Though these simulations were run without attenuation, 
{\textsc{dimo}} can accommodate these 
by an additional loss term in (\ref{Eq.Vsn}), which becomes 
\begin{equation}
  \dot{V}_{sn} 
  =
  - \int_{\cV_{\ZZ}}\bfJ(\bfr, t) \cdot \bfF_{s,-n} (\bfr) \, \rmd^{3} \bfr   
  - \sum_{m}\Omega_{s,m}I_{s,n-m}
  - \alpha_{s n} V_{s n} ,
  \label{Eq.Vsn2}
\end{equation}
with a localized positive time-decay rate $\alpha_{s n} = \alpha'(z)  v_\rmg$ for $z$ in cell $n$, 
with $v_\rmg$ the group velocity at the tube operating frequency. 
Runs with attenuation confirm the agreement between all three models.

\section{Conclusion}

We have first shown that the spatial envelope modulation assumed in
CoWAMs, which made frequency models efficient for longitudinal,
one-dimensional simulations, is inconsistent with the Maxwell
equations in the general case.  This encourages to search for
alternative modelings, and the proposed discrete model is free from
this failing.  Moreover, since time-domain models are not restricted
to a prescribed family of frequencies, they are good candidates for
investigating nonlinear regimes and the appearance of unplanned
resonances, such as a backward wave and drive induced oscillations.

Our second observations compare two one-dimensional CoWAMs and a
simple time-domain model, {\textsc{dimo}}.  On the one hand, we see
that the relevant modes for a CoWAM can be reasonably predicted 
on the basis of their possible
resonance with the beam (the backward wave is found negligible in
fig.~\ref{fig:PowerCnsup}). On the other hand, we show that the
time-domain simulation reproduces well the well-tested amplification
regime where CoWAMs are reliable.

These results show the prospects opened by time-domain direct
simulation using a compact discrete model with $\beta$-transformed
basis fields, instead of a full electromagnetic model.  An improved
version of {\textsc{dimo}} is currently under development, taking
advantage of the explicit hamiltonian nature of particle-wave
dynamics.

\medskip
\section*{Acknowledgment}

The authors are pleased to thank P.~Bernardi, F.~Doveil and D.~Minenna for fruitful discussions. 
S.~Th\'eveny was supported by a CIFRE doctoral grant.




{

\end{document}